\documentclass[letterpaper,12pt]{article}
 
	\PassOptionsToPackage{table}{xcolor}

\usepackage{abstract} 

\usepackage{titlesec} 
\usepackage{authblk} 
\usepackage{datetime} 
\newdateformat{usvardate}{\monthname[\THEMONTH] \ordinal{DAY}, \THEYEAR}

\usepackage[bottom]{footmisc}
    
    \usepackage{fancyhdr}
    \fancyheadoffset{0cm}
    \providecommand{\keywords}[1]{\textbf{\textit{Keywords---}} #1}

\newcommand\wordcount{%
  \immediate\write18{texcount -utf8 -merge -sum -incbib -dir -sub=none -brief \jobname.tex | cut -d : -f 1 > 'count.txt'}%
  \input{count.txt}\ignorespaces words%
}

    \usepackage{amssymb}
    \usepackage{amsthm}
    \usepackage{newtxmath}

    \usepackage{microtype} 
    \usepackage[utf8]{inputenc}
    \usepackage{newtxtext} 
  
	\usepackage{setspace} 

	\usepackage[top=1.5in, bottom=1.5in, left=1in, right=1in]{geometry}

	\usepackage[colorinlistoftodos]{todonotes} 
	\usepackage{soul} 
    
    \usepackage{graphicx} 
    \usepackage{float} 
    
    \usepackage{printlen}

    \usepackage{longtable} 
    \usepackage{tabularx} 
    \newcolumntype{L}[1]{>{\raggedright\arraybackslash}p{#1}}
    \newcolumntype{C}[1]{>{\centering\arraybackslash}p{#1}}
    \newcolumntype{R}[1]{>{\raggedleft\arraybackslash}p{#1}}
    \usepackage{relsize} 

	\usepackage{hyperref} 

    \usepackage{chicago} 

\makeatletter
\newcommand*{\centerfloat}{%
  \parindent \z@
  \leftskip \z@ \@plus 1fil \@minus \textwidth
  \rightskip\leftskip
  \parfillskip \z@skip}
\makeatother 

\usepackage{multirow}
\usepackage{hhline}
\usepackage{gensymb}
\usepackage{booktabs}
\usepackage{changepage}
\usepackage{mdframed}
\usepackage{algorithm}
\usepackage[noend]{algpseudocode}
\usepackage[export]{adjustbox}

\begin{document}



\thispagestyle{empty} 




  \title{\vspace{-15mm}\fontsize{21pt}{10pt}\selectfont\textbf{A Map of Knowledge}}

  \author[1,2,3,*]{\large Zachary A. Pardos}
  \author[1,4,5]{\large Andrew Joo Hun Nam}
  \affil[1]{\normalsize University of California, Berkeley}
  \affil[2]{\normalsize Graduate School of Education}
  \affil[3]{\normalsize School of Information}
  \affil[4]{\normalsize Department of Electrical Engineering and Computer Science}
  \affil[5]{\normalsize Department of Economics}
  \affil[ ]{}
  \affil[*]{\normalsize Correspondence: zp@berkeley.edu}
  \renewcommand\Authands{ and }

\date{\vspace{-5ex}}

\maketitle 



\singlespacing

\begin{abstract}
\noindent Knowledge representation has gained in relevance as data from the ubiquitous digitization of behaviors amass and academia and industry seek methods to understand and reason about the information they encode. Success in this pursuit has emerged with data from natural language, where skip-grams and other linear connectionist models of distributed representation have surfaced scrutable relational structures which have also served as artifacts of anthropological interest. Natural language is, however, only a fraction of the big data deluge. Here we show that latent semantic structure, comprised of elements from digital records of our interactions, can be informed by behavioral data and that domain knowledge can be extracted from this structure through visualization and a novel mapping of the literal descriptions of elements onto this behaviorally informed representation. We use the course enrollment behaviors of 124,000 students at a public university to learn vector representations of its courses. From these behaviorally informed representations, a notable 88\% of course attribute information were recovered (e.g., department and division), as well as 40\% of course relationships constructed from prior domain knowledge and evaluated by analogy (e.g., Math 1B is to Math H1B as Physics 7B is to Physics H7B). To aid in interpretation of the learned structure, we create a semantic interpolation, translating course vectors to a bag-of-words of their respective catalog descriptions. We find that the representations learned from enrollments resolved course vectors to a level of semantic fidelity exceeding that of their catalog descriptions, depicting a vector space of high conceptual rationality. We end with a discussion of the possible mechanisms by which this knowledge structure may be informed and its implications for data science.

\end{abstract}

\setlength\parindent{.45in} \keywords{data science, distributed representation, big data, education}




\newpage
The emergence of data science \cite{blei2017science} and the application of word vector models for representation learning \cite{mikolov2013distributed,mikolov2013efficient,mikolov2013linguistic,pennington2014glove} have, together, focused attention on surfacing structure from big data in ways that are scrutable and show signs of being able to contribute to domain knowledge. These linear connectionist models, stemming from cognitive theories of distributed representation \cite{hinton1984distributed}, have been shown to encode a surprising portion of linguistic domain knowledge learned directly from text. Generalizing out of the language context, an embedding learned from data can be framed as an informational artifact, mapping elements to parts of a structure formed by the aggregate relationships implied by their positions in a series \cite{barkan2016item2vec,pardos2017imputing,ribeiro2017struc2vec}. In our study, the elements are courses appearing in the historic enrollment sequences of tens of thousands of students at a public university. Courses themselves abstractly represent knowledge, and so our embedding, constructed from sequences of course IDs, is a map of the academically taught knowledge distributed across the university. Using this embedding, we highlight the breadth of information that can be communicated by students through course selections using a model of distributed representation applied to a dataset of modest size. In addition to interrogating the model for what prior domain knowledge it has encoded, we provide opportunities for it to surface information not previously known through visualization and semantic mapping of the learned behavioral space.


\section*{Data, Models, and Optimization}

Originally conceived of for natural language, the skip-gram and continuous bag-of-words (CBOW) models embed words into a high-dimensional vector space, with model weights adjusted through backpropagation to predict word contexts across a corpus. They can be posed as a three-layer neural network, similar in objective to an autoencoder \cite{hinton2006reducing}, creating a lower dimensional representation of the input in the hidden layer by attempting to re-construct it in the output. Unlike autoencoders, skip-grams process a single input word ($w_I$) at a time (Eq. \ref{eqn:softmax}) and capture chronology by considering only \textit{c} number of words to the left and right of the input word in calculating the loss 
(Eq. \ref{eqn:lossfunc}).  

\begin{align}
p(w_O|w_I) = \dfrac{\exp(v^{'T}_{w_O} v_{w_I})}{\sum_j \exp(v^{'T}_{w_j} v_{w_I})}
\label{eqn:softmax}
\end{align}

\begin{align}
loss =-\sum\limits_{s \in S} \dfrac{1}{T}\sum\limits^{T}_{t=1} \sum\limits_{-c \leq j \leq c, j \neq 0} \log p(w_{t+j}|w_t)
\label{eqn:lossfunc}
\end{align}

The optimized objective of the model is to increase the probability of predicting the words in context ($w_{t+j}$) given the input word in If the one-hot input layer of the model, $v_{w_I}$, were directly connected to the one-hot output layer, $w_O$, forming a multinomial logistic regression, the coefficients would simply be the distribution of target courses across all the input course’s contexts. The insertion of a hidden layer of a lower dimensionality than the number of total unique courses adds a layer of shared featurization of the courses enabling regularities to form. The input-to-hidden-layer edge weights, $w_I$, after training, yield the continuous vector representations of the courses, the collection of which is an embedding.

We used student enrollment data from UC Berkeley which spanned from Fall 2008 through Spring 2016 for a total of 23 semesters, including summer sessions, with 2,129,810 class enrollments made by 124,203 anonymized undergraduate students in 163 degree programs. Considering courses that undergraduates enrolled in, graduate courses included, there were 7,997 unique lecture courses across 197 subject areas\footnote{Subject is the most granular category of academic unit at UC Berkeley, followed by Department, Division, and College. Schools are standalone units but will be included in analyses as Subjects and Divisions.}. We encoded each course taken by a student as a one-hot “word,” allowing an undergraduate career to be represented as a sequence, $S$, of one-hots, serializing courses taken at the same time by randomizing their within-semester order. Every occurrence of a course in every student’s enrollment sequence represents a training instance, with the prediction targets being the courses in the sequence prior to and after the occurrence within a set window size. The lack of non-linear activations in this model, unlike a deep net, imparts the embedding with the properties of a vector space, allowing for arithmetic and scalar manipulation of its continuous vector representations. 
    
This collection of vectors, and the relationships they represent, is the component of interest, as opposed to the model’s predictions. It is necessary, therefore, to tune hyperparameters of the model to maximize the validity of the relationships it encodes as opposed to its predictive accuracy. In word representation learning, a sampling of domain knowledge in the broad categories of semantic and syntactic word relationships are hand defined and serve as the set of ground truth relationships with which the embedding can be validated against \cite{mikolov2013distributed}. Given our novel application to university enrollment data, one challenge was to find analogous sources of validation. We sought course relationship types which involved many departments on campus and which were as objective as possible in nature. Credit-equivalent sets of courses emerged as one such source of validation. These 128 sets consisted of 250 courses across 48 subjects, with the courses within a set declared by the Registrar’s Office as credit equivalent due to their high overlap in curriculum. We produced 381 credit-equivalent course pairs to serve as a validation set, permuted from the sets. Courses with listings in multiple departments, with distinct course IDs in those departments, served as an additional source of similarity validation with 1,472 cross-listed pairs produced from 443 cross-listed sets.

We conducted a random hyperparameter search of model topology (skip-gram vs. CBOW), window size (1 to 32), vector size (2 to 300), and three other hyperparameters. Four hundred models were trained and evaluated against 80\% of both validation sets. The nearest neighbor rank of one course in the validation pair to the other based on cosine similarity was calculated, using the median rank across pairs in a validation set as the error metric for that set (performed both ways for each pair due to asymmetry in rank). This was comparable to maximizing the relative similarity of synonymous words in the training of a language model. The best performing models were evaluated on the remaining 20\% of each validation set. A skip-gram model (vector size=229, window=8, and negative sampling=15, hierarchical softmax=0, down-sampling threshold= 7.356e-4) performed best in minimizing the combined ranks of the two 20\% sets. Selecting a model that exhibited generalizability within the equivalency and cross-listing task was important, as it would be used in subsequent analyses to generalize to completely different tasks \cite{pan2010survey}. With a learned embedding in hand, optimized using relationships between a wide swath of courses across subjects, we proceeded with scrutinizing the embedding for other forms of pedagogical regularity.

\section*{Analogy Validation}

We first evaluated the degree to which the embedding encoded five different course relationship types. The relationships between courses and their \textit{honors} version and between courses and their \textit{online} counterpart were defined from superficial course number prefixes, while pairs of \textit{sequence}, \textit{mathematical rigor}, and \textit{topical} relationships were defined using first-hand institutional prior knowledge. 

Sequence relationships were between courses prescribed to be taken in adjacent semesters in order. Many students take \textit{Mathematics} 1A and then 1B the following semester. \textit{Physics} 7A and 7B follow the same pattern, which together can form the analogical relationship, “\textit{Mathematics} 1A is to \textit{Mathematics} 1B as \textit{Physics} 7A is to \textit{Physics} 7B,” represented in vector arithmetic form as, “vec[\textit{Mathematics} 1B] - vec[\textit{Mathematics} 1A] + vec[\textit{Physics} 7A] is most cosine similar to $\rightarrow$ vec[\textit{Physics} 7B]” seen in Table \ref{table:analogies} and visualized, in part, with PCA in Fig. \ref{fig:analogy_phys}. In this approach, the representation of \textit{Mathematics} 1A is removed from \textit{Mathematics} 1B, leaving the vector offset representing the concept \cite{fodor1988connectionism,hinton1986learning} of sequence. This sequence vector is added to \textit{Physics} 7A vector, intending to yield a vector nearest to the \textit{Physics} 7B vector. The lower the nearest neighbor rank of the target course, the better the model has captured this relationship from isomorphisms in enrollment behavior. The analogy completion is only considered to have succeeded if the nearest neighbor (out of 7,996) is the anticipated target course.

We similarly isolated mathematical rigor in courses that shared content but utilized varying degrees of math. For example, while \textit{Economics} 140 and 141 both cover econometrics, 140 approaches it with a greater focus on principles with scalar operations whereas 141 uses rigorous proofs with linear algebra and probability theory. The final relationship type we coded was topical similarity between courses offered in two or three different subjects; \textit{Statistics} 155 and \textit{Economics} C110, for example, both cover game theory. The course relationship types are listed in Table \ref{table:analogies} in decreasing order of the prior domain knowledge expected to be held by students. Online and Honors courses are easily knowable from the coding syntax of the course number in the catalog. Sequences and rigor relationships, however, do not have consistent coding, but are communicated both formally by course descriptions and degree programs, and colloquially by peers and advisers. Sequences are often identifiable through suffixes (B usually follows A), but are sometimes less obvious, such as \textit{Korean} 111 following \textit{Korean} 102. Likewise, mathematical rigor (when not also an honors relationship) requires significant domain knowledge of the subject. Cross-subject topical relationships are the most difficult for students to know, requiring familiarity with the course offerings of two or three different subject areas. 

The accuracy of the course embedding in completing all 2,256 analogies generated from permutations of the 77 relationship pairs was 40\%, rivaling the 61\% seen in syntactic and semantic validations of word embeddings of Mikolov et al. \cite{mikolov2013distributed} which were trained on a dataset three orders of magnitude larger (1B words vs. 3.7M enrollments) with three times the average number of observations of each element (1,400 per word vs. 462 per course). There are no results of greater similarity to compare to as this is the first-time representations learned from behavior have been validated against propositions from domain knowledge.

\begin{table*}
\resizebox{\textwidth}{!}{%
\begin{tabular}{|c|l|}
\hline
\textbf{Relationship} & \textbf{Results} (examples) \\
\hline
Honors & \textit{Mathematics} H1B - \textit{Mathematics} 1B + \textit{Physics} 7B $\rightarrow$ \textit{Physics} H7B \\
\hline
\multirow{2}{*}{Online} & \textit{African American Studies} W111 - \textit{African American Studies} 111 + \textit{Engineering} 7\\
& $\rightarrow$ \textit{Engineering} W7 \\
\hline
Sequence & \textit{Mathematics} 1B - \textit{Mathematics} 1A + \textit{Physics} 7A $\rightarrow$ \textit{Physics} 7B \\
\hline
Mathematical Rigor & \textit{Mathematics} H1B - \textit{Mathematics} 1B + \textit{Economics} 140 $\rightarrow$ \textit{Economics} 141 \\
\hline
\multirow{4}{*}{Topical (with 2 subjects)} & \textit{Economics} C110 (game theory) - \textit{Statistics} 155 (game theory) + \textit{Statistics} 151A (linear modeling)\\
& $\rightarrow$ \textit{Economics} 141 (linear modeling) \\
\hhline{~-}
& \textit{Psychology} 102 (computing) - \textit{Psychology} 1 (introductory) + \textit{Statistics} 134 (introductory)\\
& $\rightarrow$ \textit{Statistics} H194A (honors seminar) [intended course was \textit{Statistics} 133 (computing), rank 8]\\
\hline
\multirow{5}{*}{Topical (with 3 subjects)} & \textit{Computer Science} 189 (machine learning) - \textit{Statistics} 154 (machine learning) \\
& + \textit{Statistics} 150 (random processes) $\rightarrow$ \textit{Electrical Engineering} 126 (random processes) \\
\hhline{~-}
& \textit{History of Art} 34 (Chinese art) - \textit{Chinese} 1A + \textit{Japanese} 1A\\
& $\rightarrow$ \textit{History of Art} 62 (Italian Renaissance art)\\
& [intended course was \textit{History of Art} 35 (Japanese art), rank 2]\\
\hline
\end{tabular}}
\caption{Analogy results across the six relationship types}
\label{table:analogies}
\end{table*}

We evaluated relationships between subjects in the space by querying the embedding to describe a subject as the  combination of two other subjects \cite{pardos2017school}, an analogy equation without the subtrahend. Subject vectors were created by finding the average of their respective course vectors (i.e. centroid). Expectations for these combinations were not pre-defined, as the purpose of this experiment was exploratory, presenting the results for evaluation based on their face validity. These results\footnote{A full list of all pairwise composition results can be found in extended SI.} (Table \ref{table:subjectCompositions}) suggest that there are regularities encoded not only at the micro level of the embedding, shown in the course analogies, but also more globally, as demonstrated by conceptually rational arithmetic closure at the subject level.

\begin{table*}
\centering
\begin{tabular}{@{}rll@{}}
\multicolumn{3}{c}{\textbf{Subject Compositions}} \\
\midrule
\textit{Earth \& Planetary Science} + \textit{Physics} & $\rightarrow$ & \textit{Astronomy} \\
\textit{Asian Studies} + \textit{Religious Studies} & $\rightarrow$ & \textit{Buddhist Studies} \\
\textit{Asian Studies} + \textit{Classics} & $\rightarrow$ & \textit{East Asian Languages} \\
\textit{Business Admin} + \textit{Statistics} & $\rightarrow$ & \textit{Economics} \\
\textit{Art Practice} + \textit{History} & $\rightarrow$ & \textit{History of Art} \\
\textit{Business Admin} + \textit{Computer Science} & $\rightarrow$ & \textit{Information} \\
\textit{Rhetoric} + \textit{Political Science} & $\rightarrow$ & \textit{Legal Studies} \\
\textit{Health \& Medical Sciences} + \textit{Mathematics} & $\rightarrow$ & \textit{Molecular \& Cell Biology} \\
\textit{Philosophy} + \textit{Mathematics} & $\rightarrow$ & \textit{Physics} \\
\textit{Demography} + \textit{Mathematics} & $\rightarrow$ & \textit{Statistics} \\ \bottomrule
\end{tabular}
\caption{Subject composition results}
\label{table:subjectCompositions}
\end{table*}


\section*{Visual Mapping}

\begin{figure*}
\centering
\includegraphics[width=17.8cm,height=20.18cm]{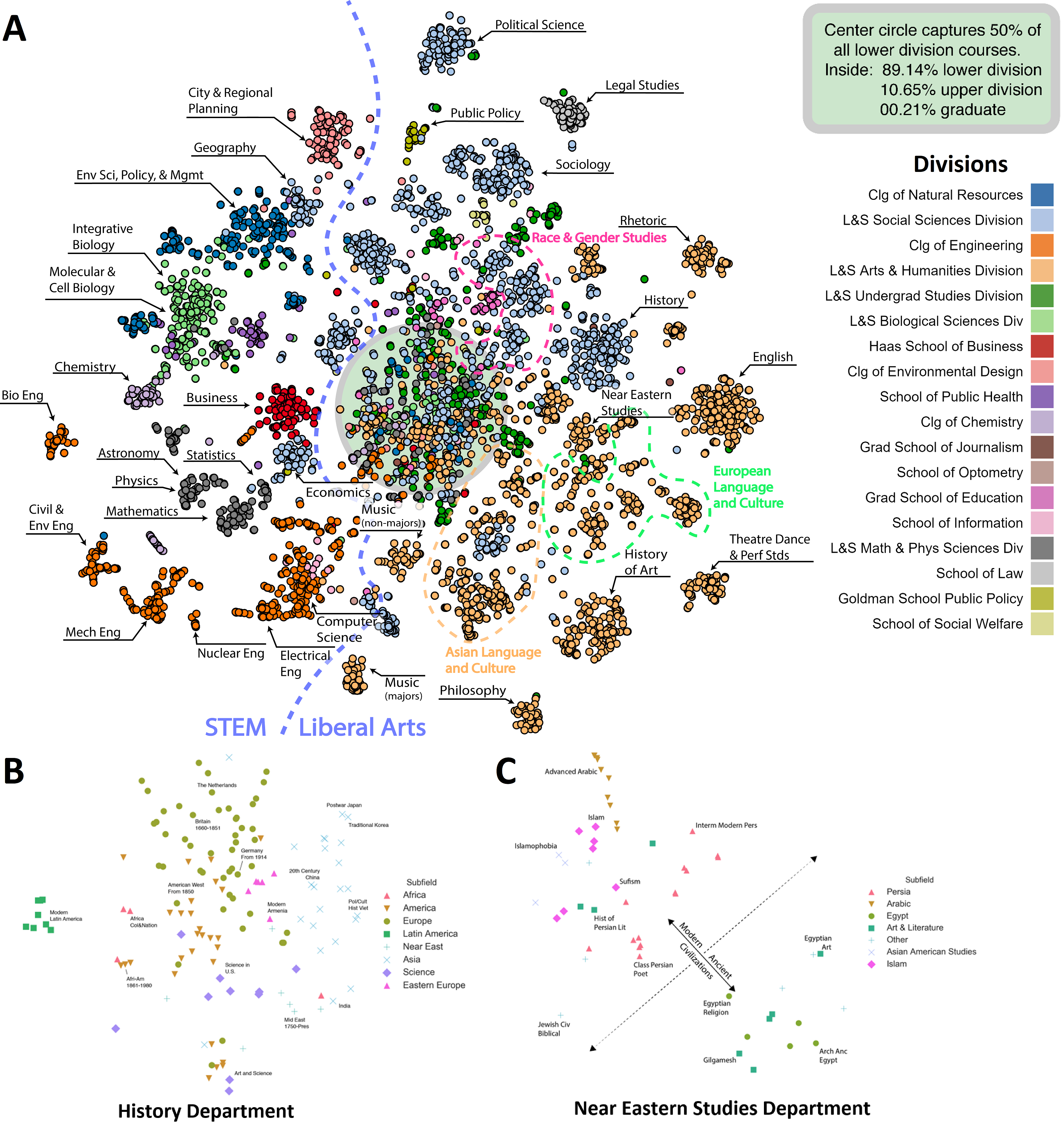}
\caption{t-SNE 2-d projection of (A) all course vectors with zoom-ins of the Departments of (B) \textit{History} and (C) \textit{Near Eastern Studies}.}
\label{fig:tsne}
\end{figure*}

We visualized the course embedding to surface the primary factors which dictate vector proximity in the space, using Barnes-Hut t-SNE\footnote{t-SNE default parameters were used: perplexity 30, theta 0.5, and initial PCA to 50 dimensions.} \cite{vandermaaten2014accelerating} for dimensionality reduction. This allowed for observation of micro, meso, and macro scale relationships not hypothesized and produced a never before seen view of the university and the relationships between its disciplines. Each data point in Fig. \ref{fig:tsne}A is a course, colored by the division it belongs to, with labels added for subject groupings. t-SNE prioritizes the retention of local structure from the high-dimensional space in its manifold projection to the two-dimensional space, thus excelling at depicting local structure in an embedding.

At the micro-level, the visualization reveals salient conceptual relationships between individual courses. Zooming into the History cluster, the courses organize roughly into a rotated map of the globe (Fig. \ref{fig:tsne}B). Starting at the top right are the east Asian countries: Japan and Korea with China to their west. Below them are southeast Asian countries such as Vietnam and India to its west. Towards the west, we find eastern Europe, western Europe, and finally United States. Though some clusters do not adhere perfectly, this geographical layout can be explained by the specialization in a time and a place among historians, and thus their students, with some interest in adjacent regions but less emphasis on cross-cutting global themes. Where the norms of the \textit{History} department placed courses geographically, Near Eastern Studies separates them temporally, with a boundary between courses covering modern and ancient civilizations (Fig. \ref{fig:tsne}C). We find that ancient literature, religions, and societies such as Egypt, map towards the lower right whereas modern languages and religions such as Arabic and Islam, populate the top left, representing the discipline’s bi-modal foci.

Logical meso-level relationships can also be seen, with \textit{Statistics} situated between \textit{Mathematics} and \textit{Economics} and \textit{Physics} between \textit{Mathematics} and \textit{Astronomy} (Fig. \ref{fig:tsne}A). An interesting path begins in \textit{Chemistry}, traversing through \textit{Molecular \& Cell Biology}, \textit{Integrative Biology}, \textit{Environmental Science \& Policy Management}, \textit{Geography}, \textit{City \& Regional Planning}, and terminating at \textit{Architecture}. The subjects progress with conceptual coherence between neighbors such that, though \textit{Chemistry} and \textit{Architecture} may have little in common, the relationship between each intermediary subject is logical. This adjacency of disciplines naturally bears resemblance to relationships seen in the broader study of academic research diffusion \cite{boyack2005mapping,rosvall2008maps}. While the majority of courses grouped by subject, interdisciplinary groupings were observed in the thematic areas of Race \& Gender Studies, European Language \& Culture, and Asian Language \& Culture (Fig. \ref{fig:tsne}A, \ref{fig:tsne_meso_rgs}, and \ref{fig:tsne_meso_alc}).

A noticeable characteristic surfaced in the visualization is the unstructured cloud of largely Lower Division level courses near the origin, contrasted against the more structured clusters of Upper Division courses outside it. Berkeley classifies Lower Division courses as part of the introductory sequences to an academic discipline often taken by prospective students of the associated program or to fulfill Berkeley’s mandatory breadth requirements. As a set of exploratory courses, Lower Division courses are expected to generate higher degree of variance in the enrollments contexts in which they appear, whereas Upper Division courses assume certain prior knowledge of their subjects and are often taken by students after the lower division courses, thereby embedding further away from the center due to demonstrating lower entropy.

Finally, at the macro level, a bisection of the entire map divides subjects considered to be Science, Technology, Engineering, and Math (STEM\footnote{Immigration, U. S., and Customs Enforcement. "STEM-designated degree program list." (2016).}) on the left side from Liberal Arts subjects on the right. Courses offered by the College of Engineering reside close to the bottom left quadrant, natural sciences to the upper left, social sciences in the upper right, and arts \& humanities in the bottom right. Departments under the Social Sciences Division are largely found in the Liberal Arts hemisphere with the exceptions of \textit{Psychology} and \textit{Economics}, both of which have highly statistical facets. Though the STEM classification of courses in our embedding is not new knowledge, it demonstrates that the embedding can capture information not likely fully known by any of the individuals whose actions it was produced from. This observation also underscores the impressive ability of t-SNE to render a single projection with conceptual coherence retained at several levels of scale.

The salience of clustering by subject in the visualization begged the question of what other course attribute information was encoded in the embedding. To quantify this, we trained multinomial logistic regression models, using course vectors as the input to regress to six different categorical attributes detailed by the Registrar’s Office and our enrollment metadata. These models performed well in predicting the attribute values of a held-out test set of course vectors, with the subject of a course predicted with 84.19\% accuracy based on its vector compared to 3.01\% when predicting using the most common subject. Overall, attribute values were predicted with 87.95\% accuracy using the embedding compared to 30.63\% by majority class (Table \ref{table:tagpredictions}).

\begin{table}
\centering
\begin{tabular}{lrrr} 
\textbf{Attribute} & \textbf{Unique Values} & \textbf{Majority} & \textbf{Logistic} \\
\midrule
Subject & 197 & 3.01\% & 84.19\% \\
Department & 81 & 5.01\% & 87.06\% \\
Division & 20 & 27.15\% & 84.92\% \\
College & 16 & 64.61\% & 94.60\% \\
Course Level & 3 & 57.11\% & 91.06\% \\
Modal Major & 114 & 26.88\% & 85.86\% \\
\hline
& \textit{Average} & 30.63\% & 87.95\% \\
\end{tabular}
\caption{Results of predicting attributes from course vectors}
\label{table:tagpredictions}
\end{table}

\section*{Semantic Mapping}

While rich in structure, a learned embedding lacks interpretability without added semantics. In the previous sections, we used the academic unit meta information of courses to add semantics and interpret their compositionality with respect to those semantics with coloring in the visualization and through projecting to subject centroids. In the close-up of \textit{History} and \textit{Near Eastern Studies} (Fig. \ref{fig:tsne}B and \ref{fig:tsne}C), experts\footnote{Identified in the paper's Acknowledgements.} were consulted to aid in the interpretation of the locations of courses in the plots. In this section, we instead use course descriptions to automatically provide insights about the space. This queriable semantic mapping of a vector space can itself be seen as an expert, whose epistemic expertise is defined by its ability to justify its knowledge through propositions in the domain \cite{weinstein1993expert} in contrast to a deep net, whose expertise is defined by its ability to perform \cite{krizhevsky2012imagenet,mnih2015human,silver2016mastering,moravvcik2017deepstack}, both with distributed representation at the core of their generalizing principle. To do this, we trained a multinomial logistic regression mapping course vectors to their bag-of-words course descriptions sourced from the university course catalog. This was a machine translation, not between languages \cite{mikolov2013exploiting} but between a course representation space formed from behaviors and a semantic space constructed from instructors’ descriptions of the knowledge imparted in each course. This mapping allowed arbitrary vectors in the space to be semantically described using keywords, those descriptions regularized by way of their regression from the embedding. To control the level of specificity of the words outputted by the model, we introduced a bias parameter (Eq. \ref{eqn:tfidf}). A higher bias would result in words that could be considered discipline-specific jargon, while a lower bias would produce descriptions using more general and accessible words. While we initially applied tf-idf \cite{salton1986introduction}, the brevity of course descriptions usually yielded at most one instance of each word in a description, effectively nullifying the term-frequency weight component and reducing tf-idf to only idf. Experimentally, we found that treating the entire collection of descriptions as one document and exponentiating the raw frequency to a negative number yielded desirable contrast in word specificity.

\begin{align}
tf\text{-}bias = \left(\dfrac{number\ of\ occurrences\ of\ word}{total\ word\ count}\right)^{-bias}
\label{eqn:tfidf}
\end{align}

We first ran the subject vectors through the semantic model to see whether a simple average of course vectors captured the core of subjects. Table \ref{table:subjdesc}\footnote{Semantic model descriptions of all subjects at the University can be found in SI datasets.} compares biases 0.5 and 1 across three exemplar subjects. A bias of 0.5 preferred broader words such as “Algorithms”, “Markets”, and “Society” in \textit{Computer Science}, \textit{Economics}, and \textit{Sociology}, respectively, while a bias of 1 surfaced “Robotics”, “Game Theory”, and “Comparative perspective” in those subjects. Particularly frequent descriptions appeared in both bias lists, such as “Computer”, “Industrial organization”, and “Inequality.”

\begin{table*}
\resizebox{\textwidth}{!}{%
\begin{tabular}{cccccc}
\multicolumn{2}{c}{\textbf{\textit{Computer Science}}} &  \multicolumn{2}{c}{\textbf{\textit{Economics}}} &  \multicolumn{2}{c}{\textbf{\textit{Sociology}}} \\
0.5 & 1 & 0.5 & 1 & 0.5 & 1 \\
\midrule
Computer & Algorithms & Economic & Economic & Sociological & Sociological \\
\rowcolor[HTML]{EFEFEF} 
Design & Computer & Theory & Industrial organization & Social & Inequality \\
Algorithms & Computer science & Analysis & Size & Inequality & Social change \\
\rowcolor[HTML]{EFEFEF} 
Techniques & Program language & Determinants & Linear regression models & Social change & Social \\
Models & Implementation & Policy & Pricing & Society & Hypotheses \\
\rowcolor[HTML]{EFEFEF} 
Control & Codes & Markets & Boom & Theory & Trends \\
Data & Machine & Development & Econometric & Institutions & Thought \\
\rowcolor[HTML]{EFEFEF} 
Applications & Privacy & Pricing & Income & Thought & Dominant \\
 & Artificial & Industrial & Game & & Comparative \\
\multirow{-2}{*}{Structure} & intelligent & organization & theory & \multirow{-2}{*}{Trends} & perspective \\
\rowcolor[HTML]{EFEFEF} 
Project & Robotics & Size & Valuation & Within & European countries \\
\bottomrule
\end{tabular}}
\caption{Semantic model descriptions of Subject vectors using biases of 0.5 and 1.}
\label{table:subjdesc}
\end{table*}

We then asked the model to describe three subjects (\textit{Design Innovation}, \textit{Neuroscience}, and \textit{Plant Biology}) for which not a single course’s description from the subject was part of the semantic model training\footnote{Catalog descriptions of courses in these subjects were missing due to a limitation of the API used to access the catalog at the time, creating a naturally occurring opportunity for an experiment.}. \textit{Neuroscience}, for example, produced words such as “brain”, “physiology”, “sensory”, and “neuroanatomy,” words likely borrowed from other subjects in biology. \textit{Design Innovation} produced apt words such as “team”, “user”, “technology”, “interface”, and “robotics.” These descriptions, in Table \ref{table:vec2text}, demonstrated the model’s ability to interpolate semantic meaning across sparse regions of the space.

\begin{table*}
\resizebox{\textwidth}{!}{%
\begin{tabular}{ccccccc}
\textbf{\textit{Design}} & \multirow{2}{*}{\textbf{\textit{Neuroscience}}} & \textbf{\textit{Plant}} & \textbf{Origin} & \textbf{ECON 141} & \textbf{MATH H113}& \textbf{ARTHIST 32}\\
\textbf{\textit{Innovation}} &  & \textbf{\textit{Biology}} & \textbf{Vector} & - \textbf{ECON 140} &  - \textbf{MATH 113} & - \textbf{JAPAN 1A} \\
\midrule
Team & Brain & Microbial & Cultural & Variants & Enjoy & Tumuli \\
\rowcolor[HTML]{EFEFEF} 
Enable & Human brain & Molecular & History & Vector & Hidden & Seventeenth \\
User & Physiology & Preservation & World & Theorem & Hard & Newcomers \\
\rowcolor[HTML]{EFEFEF} 
Share & Neurological & Plant & Social & Mathematical & Beauty & Neolithic \\
Innovation & Sensory & Biotechnology & Development & Quadratic forms & Corresponding & Nineteenth century \\
\rowcolor[HTML]{EFEFEF} 
Perception & Biology science & Habitat & Society & Eigenvectors & Recommended & Proceed \\
Technology & Neural & Metabolic & Language & Discrete continuing & Honors & Art architecture \\
\rowcolor[HTML]{EFEFEF} 
Interface & Neuroanatomy & Genomics & Political & Integer & Rigorous & Chronological \\
 &  &  &  & Function complex &  &  \\
\multirow{-2}{*}{Robotics} & \multirow{-2}{*}{Neurophysiology} & \multirow{-2}{*}{Genetics} & \multirow{-2}{*}{Modern} & variables & \multirow{-2}{*}{Inclination} & \multirow{-2}{*}{Focus particular} \\
\rowcolor[HTML]{EFEFEF} 
 &  &  &  & Conditions &  &  \\
\rowcolor[HTML]{EFEFEF} 
\multirow{-2}{*}{Vision} & \multirow{-2}{*}{Anatomy} & \multirow{-2}{*}{Biology} & \multirow{-2}{*}{Human} & expected & \multirow{-2}{*}{Greater} & \multirow{-2}{*}{Realism} \\
\bottomrule
\end{tabular}}
\caption{Semantic model description of missing Subjects, the origin vector, and course vector differences (0.5 bias)}
\label{table:vec2text}
\end{table*}

An emergent \cite{hopfield1982neural} set of vectors from course analogies were the vector offsets between two courses with the isomorphisms in course analogies suggesting that the difference vector was itself representative of a shared distributed concept. We used the semantic model to describe these vector offsets. For instance, subtracting \textit{Japanese} 1A (‘Elementary Japanese’) from \textit{History of Art} 32 (‘Art and Architecture of Japan’) produced a vector described by the semantic model as, “tumuli”, “Neolithic”, “art-architecture”, and “realism,” words appropriate for describing art history. While we ascribed the relationship between \textit{Economics} 141 and 140 as a more mathematically "rigorous" treatment of econometrics, the semantic model succeeded in articulating more granular pedagogical differences, using words like “vectors,” “discrete-continuous,” and “conditional expectations” to accurately describe the content in 141 but not in 140 from the offset vector\footnote{This offset vector, produced by subtracting ECON 140 from ECON 141, had two courses in linear algebra, MATH 110 and MATH 113, as its nearest neighbors.}. Other words that appeared such as “quadratic forms” and “eigenvectors,” while not explicitly taught as part of the course material, are related to linear algebra, the topic only found in the more advanced offering (Table \ref{table:vec2text}). The semantic model, leveraging a rich vector space formed from behaviors, surfaced these topical differences not found in either course's catalog description:

\mdfdefinestyle{pnassigstyle}{linewidth=0.7pt,innertopmargin=6pt,innerrightmargin=6pt,innerbottommargin=6pt,innerleftmargin=6pt,font=\small}
\begin{mdframed}[style=pnassigstyle]
\begin{flushleft}
\textbf{Economics 140:} Introduction to problems of observation, estimation, and hypothesis testing in economics. This course covers the linear regression model and its \underline{application to} empirical \underline{problems in} economics.
\linebreak
\textbf{Economics 141:} Introduction to problems of observation, estimation, and hypothesis testing in economics. This course covers the \underline{statistical theory for} the linear regression model and its \underline{variants, with examples from} empirical economics.
\end{flushleft}
\end{mdframed}

The ability to describe any arbitrary vector allows for queries that have no correspondence to a particular course, but are conceptually interesting nonetheless. The origin vector could be interpreted as the center of Berkeley’s academic demography but otherwise has no educational meaning. The semantic model describes the origin with the words “cultural,” “history,” “world,” “social,” and “development” as the top five results, which may be a reasonable way to describe the liberal-arts centric campus of UC Berkeley.

\section*{Conclusions}
Visualization of the course embedding at several scales evokes images of cell-cultures in a petri dish under a microscope or a deep field view of constellations through a telescope. This paper's domain of study can be viewed analogically as elements – courses - introduced into the social system of a university with human factors serving as the forces dictating the movement of the elements and their positionality in the structure as a whole. This representational structure, illuminated by data and studied through the instrument of a learned embedding analysis, is analogous to the physical structures studied with instruments from the natural sciences and is part of a larger universe of explorable structure expanding at the speed of data collection. A question of natural concern to the developing notion of data science is whether truths can be learned from behavioral data through this lens of a representation analysis. Our study used a variety of inference types to interrogate the embedding for such truths: abductive inference to describe patterns in the visual mapping, inductive inference to define subjects by an aggregation of their courses, and deductive inference to validate analogies\footnote{Analogies can be viewed as syllogisms (e.g. All honors courses are vec[honors] from non-honors courses, course A is vec[honors] away from a non-honors course, therefore course A is an honors course)}. If truths about courses were to be defined as the instructors’ catalog descriptions, then the semantic interpolation was able to successfully surface previously unknown truths about the subjects of courses with no catalog descriptions and about topical difference between courses. It is expected that when applied to other data contexts, semantics about elements truly unknown to a domain could be revealed. The embedding encoded 40\% of relationships from prior domain knowledge and 88\% of course attributes, both of which could also be considered truths. We therefore conclude that considerable knowledge is encoded and made accessible using these methodologies, from representational structure formed by behaviors alone; with the validity of individual inferences dependent on the veracity of the regularities, known to increase with data volume. 

Broadly, the embedding may encode attributes and aggregated tacit knowledge of courses by mechanisms such as the wisdom of crowds \cite{galton1907vox,surowiecki2004wisdom}, distributed cognition \cite{hutchins1995cognition}, or the combination of expert opinions \cite{dawid1995coherent} or classifiers \cite{jacobs1991adaptive}. However, like the cultural biases reflected in word embeddings \cite{caliskan2017semantics,bolukbasi2016man}, a course embedding too has an anthropological epistemology. It is perhaps most aptly characterized as students’ perceptions of courses at the time of enrollment, influenced by peer testimonials and degree requirements (faculties’ representations of their relatedness). In this sense, the embedding, and data science itself, takes on a dual identity of aiding in the pursuit of truths on one hand and on the other, reflecting the disposition of the individuals and society whose data it is constructed from.

\newpage
\singlespacing
\bibliographystyle{references/pnas}
\bibliography{references/biblio}
    
\newpage
\section*{Supporting Information (SI)}
\setcounter{table}{0}
\setcounter{figure}{0}
\def\thealgorithm{S\arabic{algorithm}}
\renewcommand{\thetable}{S\arabic{table}}
\renewcommand{\thefigure}{S\arabic{figure}}

\subsection*{Course Data}

To distinguish between special topics courses, a set of courses that are cataloged with the same course number although the course material depends on the instructor, we appended the course identifier with the instructor’s name to distinguish them from one another. While the skip-gram model maintains robustness with sufficient data per label, it suffers from noise when there are too few data points. To reduce this type of noise, we filtered out courses that had less than 20 enrollments in the 8 years the dataset covered. We also removed generic placeholder courses for non-curricular activities such as independent research and senior theses “courses,” decreasing the unique courses in the model from 7,997 to 4,349. 

The sets of cross-listed courses, used as a validation set, were generated by aggregating non-summer courses that were listed differently, but shared the same room at the same time whenever offered during the same semester. These were courses that had different course listings (e.g. \textit{Economics} C175 and \textit{Demography} C175) but were in every other respect the same including shared lectures, discussion sections, assignments, and grading distributions (i.e. a student could enroll in either course and it would make no difference). If two courses were not cross-listed with each other, but were cross-listed with a same third course, then we considered the set of three to be cross-listed. We only counted courses as cross-listed where the courses were cross-listed every semester both courses were offered.

Course credit equivalencies, also used as a validation set, were collected from schedulebuilder.berkeley.edu (now deprecated) and manually parsed through due to the natural language wording of the equivalencies, varying conditions for credit disqualification, and partial credit disqualifications. We chose to treat all levels of equivalency (full, partial, conditional) the same, assuming all equivalencies exhibited conceptual similarity that could surface signal in the validation set. Lastly, we allowed for two courses to be considered equivalent as long as they shared a third course that was equivalent with the two. This set distinguishes itself from the cross-list sets where the courses in those sets are the same, not just equivalent. 

\subsection*{Model Architecture and Tuning}

If the one-hot input layer of the model, $w_I$, were directly connected to the one-hot output layer, $w_O$, forming a multinomial logistic regression, the coefficients would simply be the distribution of target courses across all the input course’s contexts. The insertion of a hidden layer of a lower dimensionality than the number of total unique courses adds a layer of shared featurization of the courses enabling regularities to form. The input-to-hidden-layer edge weights, $w_I$, after training, yield the continuous vector representations of the courses, the collection of which is an embedding.

We optimized across 6 model hyperparameters of the course2vec models using random search: the model architecture (skip-gram vs CBOW), window size, vector size, the use of hierarchical softmax, the use of negative sampling \& the number of noise words to draw during negative sampling, and the threshold for down-sampling higher-frequency words. Using the cross-list sets and the credit equivalency sets, we performed round-robin queries for nearest neighbors for each set, taking the median rank for each set, then taking the median rank across all the sets (Algorithm \ref{alg:validation}). Because optimizing by a different metrics would yield different models, we allowed the optimization metric to be another point of comparison.

\begin{algorithm}
\caption{Validation Score}\label{alg:validation}
\begin{algorithmic}[1]
\Procedure{Validation Score}{$validation\_set$}
	\State medians $\gets$ new list
    \For{$\forall set s \in validation\_set$}
		\State scores $\gets$ new list
        \For{$\forall course c \in s$}
    		\For{$\forall$ course $c' \in s, c \neq c'$}
          		\State scores.add(rank of $c'$ using nearest neighbor to $c$)
        	\EndFor
    	\EndFor
    	\State set\_score $\gets$ median(scores)
		\State medians.add(set\_score)
    \EndFor
  \State $validation\_score \gets median(medians)$
\State \textbf{return} validation\_score
\EndProcedure
\end{algorithmic}
\end{algorithm}

Cross-listed courses not only provided a validation set, but also a new way to organize the data during preprocessing. Two cross-listed courses are only nominally different and therefore could be substituted with the other without any loss in its logical representation of a student’s enrollment history. Because course2vec maps a course within context of its neighbors, even the nominal difference of being listed under a different subject placed cross-listed course vectors far away from each other despite their ontological equivalences. For example, while a logical mapping would place \textit{Economics} C110 and \textit{Political Science} C135 on the same point, a context-based embedding would place the former in the cluster of \textit{Economics} courses while the latter would map to the \textit{Political Science} cluster. Students majoring in economics tended to enroll in the \textit{Economics} offering, making its neighboring courses other \textit{Economics} courses. Alternatively, collapsing the cross-listed courses would force these courses to share all contexts, bringing \textit{Economics} and \textit{Political Science} clusters closer together as well. We tested whether this would improve the embedding by collapsing none, half, and all of cross-listed courses and comparing the performance of the resulting models.

A model with no collapsed cross-listed courses would have the full set of cross-listed courses represented in its score while a model with all collapsed would have none (80\% collapse could not be optimized by cross-lists). To compare across the models optimized by different metrics and different cross-list collapse proportions, we held out 20\% of the validation sets during their optimization since they could not be compared using their validation scores. Since this would result in models with 80\% of cross-lists collapsed, we used the exact hyperparameters of the best model with 40\% and 80\% collapse to generate a model with 100\% collapse for the cross-list and equivalence optimizations respectively. 

\begin{figure}[tbhp!]
\centering
\includegraphics[width=.7\linewidth]{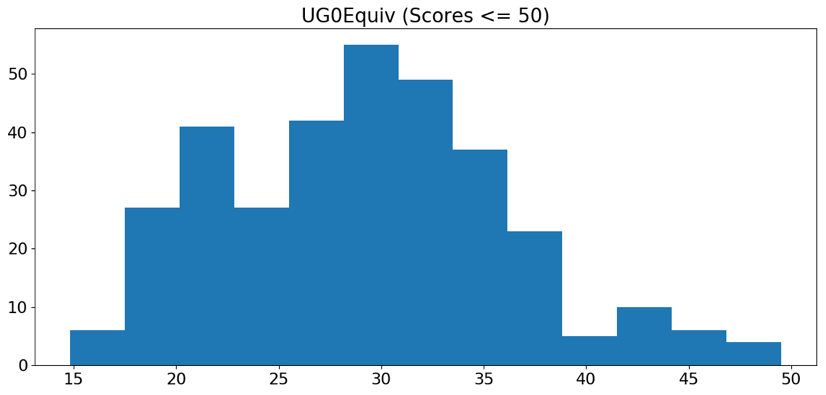}
\caption{Distribution of model (equivalence) scores. Scores > 50 omitted due to high skew}
\label{fig:scores}
\end{figure}

We ran each class (optimization metric and cross-list collapse) of optimizations separately generating 400 models per class, which ran for approximately 24 hours on a high-performance computer.  Taking the best models of each class and comparing their test set scores, we found that optimizing to equivalence set score produced superior results (Table \ref{table:scores}). Even when taking the average of both scores, the score in equivalencies dominated the validation score such that the resultant model of taking the average would yield the same model as optimizing on only the equivalency sets (using a seeded random search). Moreover, when optimizing on cross-list sets, the validation score for equivalencies would be penalized significantly more than the validation score for cross-lists would be when optimizing on the equivalency sets. Consequently, in our paper, we used equivalency sets as the primary validation set to select our best model. Among the remaining models, the models with no collapsed cross-listed performed best, leaving the uncollapsed model optimized by equivalence sets the final model. Fig. \ref{fig:scores} shows the distribution of scores when optimizing to the equivalence set from which the best model was selected.

\begin{table*}
\resizebox{\textwidth}{!}{%
\begin{tabular}
{|p{2.2cm}|p{2.2cm}|p{2cm}|p{2.3cm}|p{1.8cm}|p{2cm}|p{2.3cm}|p{1.2cm}|}
\hline
\textbf{Optimization Metric} & \textbf{Proportion of Collapsed Cross-listed Courses} & \textbf{Cross-lists Validation Score} & \textbf{Equivalencies Validation Score} & \textbf{Overall Validation Score} & \textbf{Cross-lists Test Score} & \textbf{Equivalencies Test Score} & \textbf{Overall Test Score} \\ \hline
Cross-lists & 0\% & 12 & 47.5 & 29.75 & 19.5 & 181 & 100.25 \\ \hline
Cross-lists & 40\% & 10 & 48.5 & 29.25 & 16.5 & 303.5 & 160 \\ \hline
Cross-lists & 100\% & \cellcolor[HTML]{000000} & 42 & 42 & \cellcolor[HTML]{000000} & 303.5 & 303.5 \\ \hline
Equivalence & 0\% & 23.5 & 17 & 20.25 & 28 & 31.5 & 29.75 \\ \hline
Equivalence & 40\% & 16.5 & 16.5 & 16.5 & 33 & 33 & 33 \\ \hline
Equivalence & 80\% & \cellcolor[HTML]{000000} & 16 & 16 & 59 & 39 & 49 \\ \hline
Equivalence & 100\% & \cellcolor[HTML]{000000} & 16 & 16 & \cellcolor[HTML]{000000} & 38 & 38 \\ \hline
\end{tabular}}
\caption{Best model scores by optimization and proportion of collapsed cross-listed courses}
\label{table:scores}
\end{table*}

\begin{figure}[tbhp!]
\centering
\includegraphics[width=.7\linewidth]{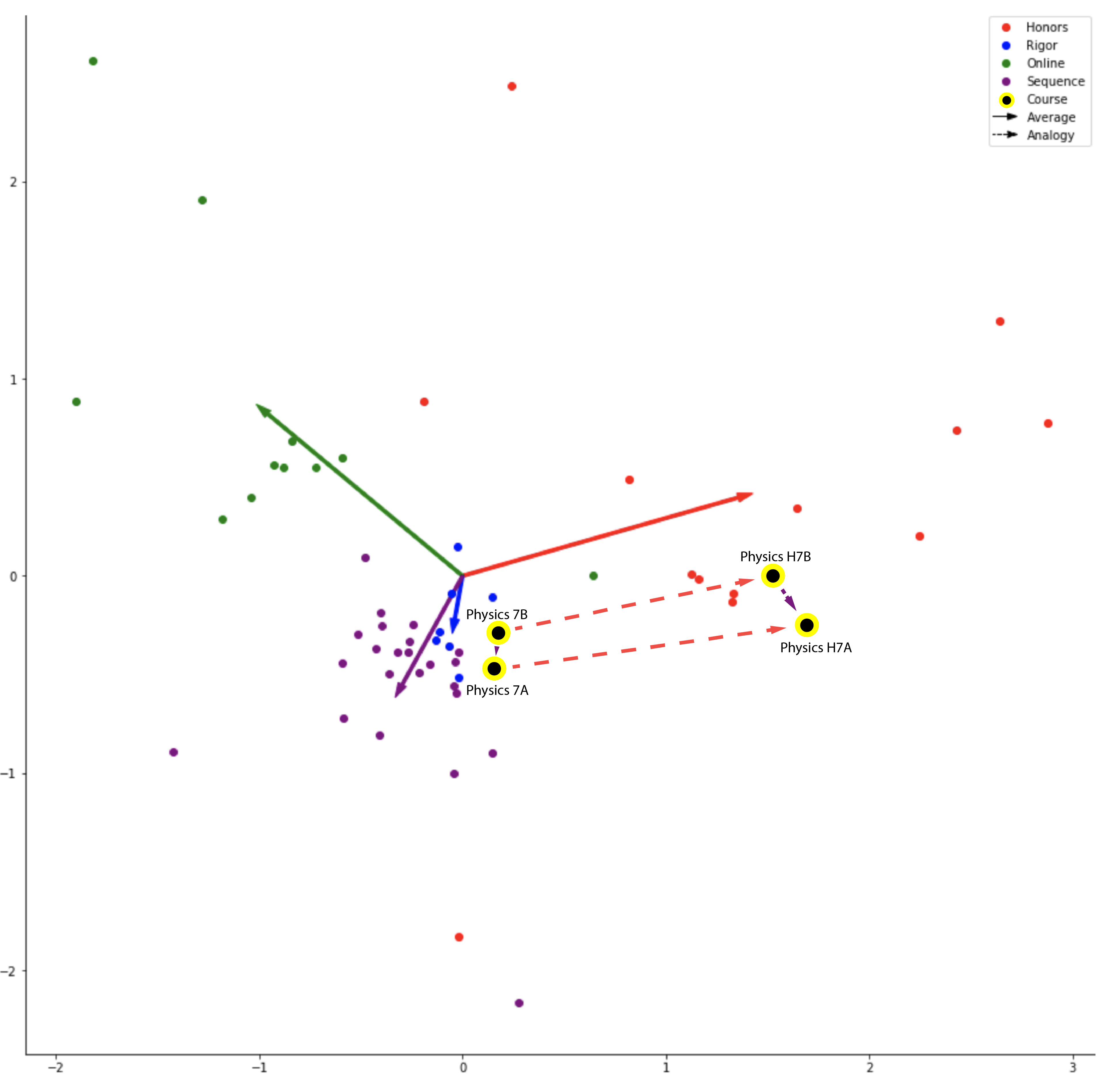}
\caption{PCA of vector offsets with Sequence and Honors constellation using \textit{Physics} courses}
\label{fig:analogy_phys}
\end{figure}

\subsection*{Analogy Validation}

In creating sequence, mathematical rigor, and topical relationships, we manually identified course pairs with such features using prior knowledge of UC Berkeley’s courses. Honors and online relationships were marked with an ‘H’ or ‘W’ prefix and therefore were generated automatically.

For all but the topical relationships, since any two pairs of courses within a relationship set were expected to share the same regularity, we tested the analogies in round-robin fashion (e.g. comparing one sequence pair with every other every other sequence pair and finding its rank for each). Moreover, by shuffling the order of courses in each analogy, we generated a total of roughly (removing overlaps)  analogy equations where N = the number of pairs in the relationship set. With 23 sequence pairs, 18 mathematical rigor pairs, 14 honors pairs, and 12 online pairs, we generated 1008, 576, 364, and 264 analogy equations respectively. Since topical relationships required two comparable course pairs and therefore lacked the fungibility of the other relationship types, we generated $4 \times N$ analogies using N = 11 quadruples for a total of 44 analogy equations.

To visualize the regularities formed by analogous structures, we projected vector offsets using PCA and applied the transform to the vectors of \textit{Physics} 7A, \textit{Physics} H7A,  \textit{Physics} 7B, and \textit{Physics} H7B to represent the Sequence and Honors relationships, creating an imperfect formation of an analogy constellation (Fig. \ref{fig:analogy_phys}).

\subsection*{The vector to text model (Semantic Mapping)}

We collected descriptions of courses from Berkeley’s Course API and concatenated them with the course titles. We also removed stop-words (e.g. the), stemmed words using the snowball algorithm, and used iterative bigram phrase detection before collecting the words into bags-of-words vectors. Because some courses shared the exact same titles and descriptions, to reduce false positives during phrase detection, we only allowed one instance of duplicates to be traversed. To remove vague words and words related to course logistics, we filtered words across 4 different metrics, taking 100 words in each and hand-selected from the set (Table \ref{table:v2tpreprocess}). For words that could be meaningful in certain contexts, we chose to remain conservative and include the words within the vocabulary. For example, the phrase ‘web site’ may indicate that a course is taught through an online medium, but could be contextually relevant in subject areas such as design, media, and information.

\begin{table*}
\resizebox{\textwidth}{!}{%
\begin{tabular}{|p{1.3cm}|p{2cm}|p{2.5cm}|p{2cm}|p{2.2cm}|p{2.1cm}|}
\hline
\textbf{Target} & \textbf{Sorting Metric} & \textbf{Reason} & \textbf{Top Removed} & \textbf{Top Kept} & \textbf{Total Removed} \\ \hline
Phrases & Number of occurrences & Common phrases are more likely to reflect logistics & Freshman sophomore seminar & Case study & 68 phrases \\ \hline
Words & Number of occurrences & Common words are more likely to be vague & Course & Development & 51 words \\ \hline
Words & Number of subjects they appear in & Breadth of words suggest vagueness and logistics & Covered & Current & 76 words \\ \hline
\end{tabular}}
\caption{Rules for removing descriptions from semantic model training corpus}
\label{table:v2tpreprocess}
\end{table*}

Using the final descriptions, we trained multinomial regression models where the course vectors were used as input features and corresponding course descriptions as bag-of-words multi-hot output vectors, scaled by their tf-bias weights (Table \ref{table:subjdesc}). We used the same training parameters as the tag prediction model, but left epochs as a hyperparameter rather than using early-stopping which often failed to trigger. 

We performed a grid search to roughly optimize bias and epoch parameters. Descriptions being largely qualitative, we inspected a sample subset of models to judge the words generated by the models. Empirically and matching intuition, lower bias and epoch count produced more vague words whereas higher bias and epoch count produced more specific words.

\clearpage
\clearpage

\begin{figure*}[tbhp!]
\centering
\includegraphics[width=.7\linewidth]{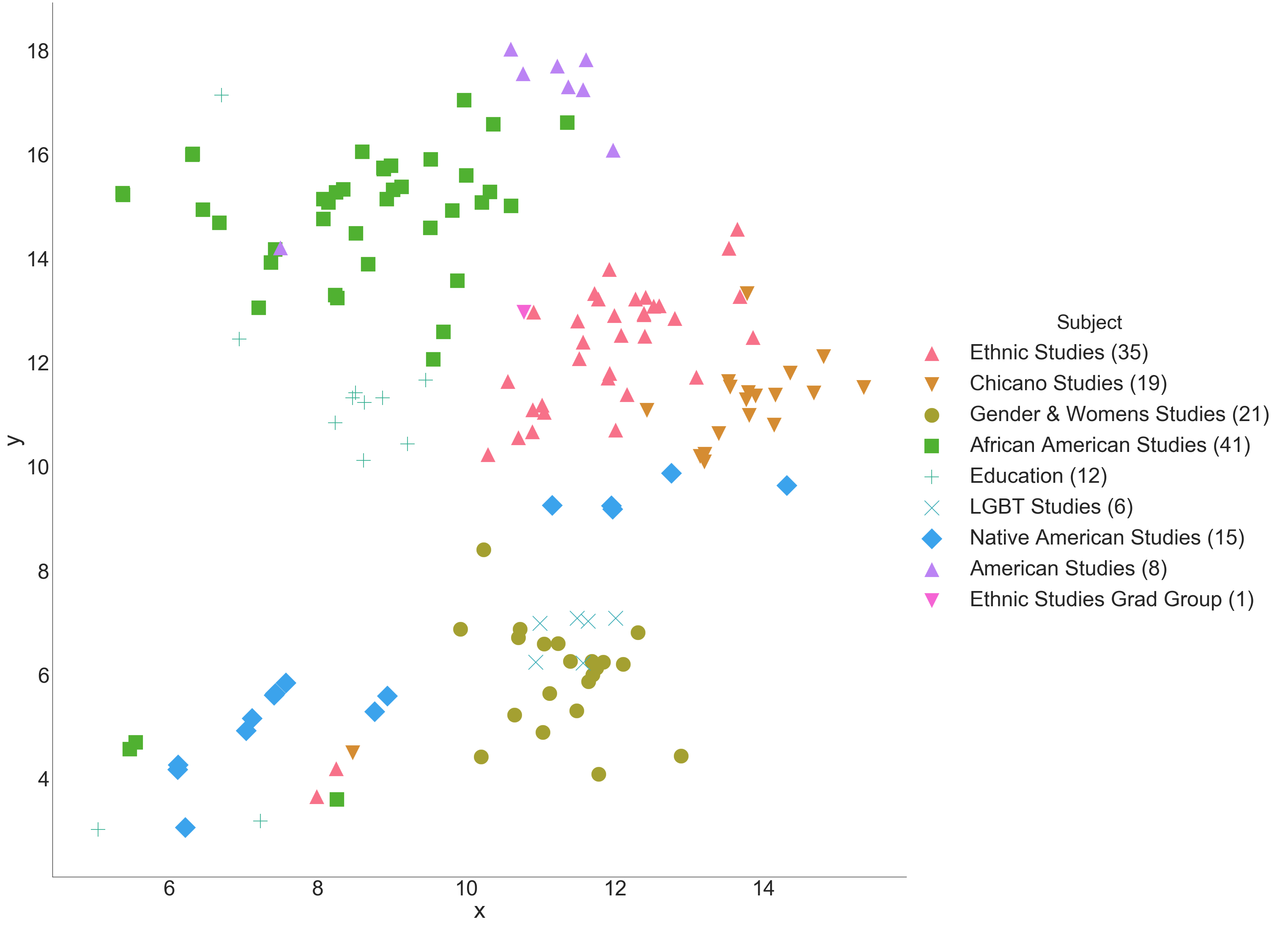}
\caption{Zoom-in of Race \& Gender Studies cluster}
\label{fig:tsne_meso_rgs}
\end{figure*}

\begin{figure*}[tbhp!]
\centering
\includegraphics[width=.7\linewidth]{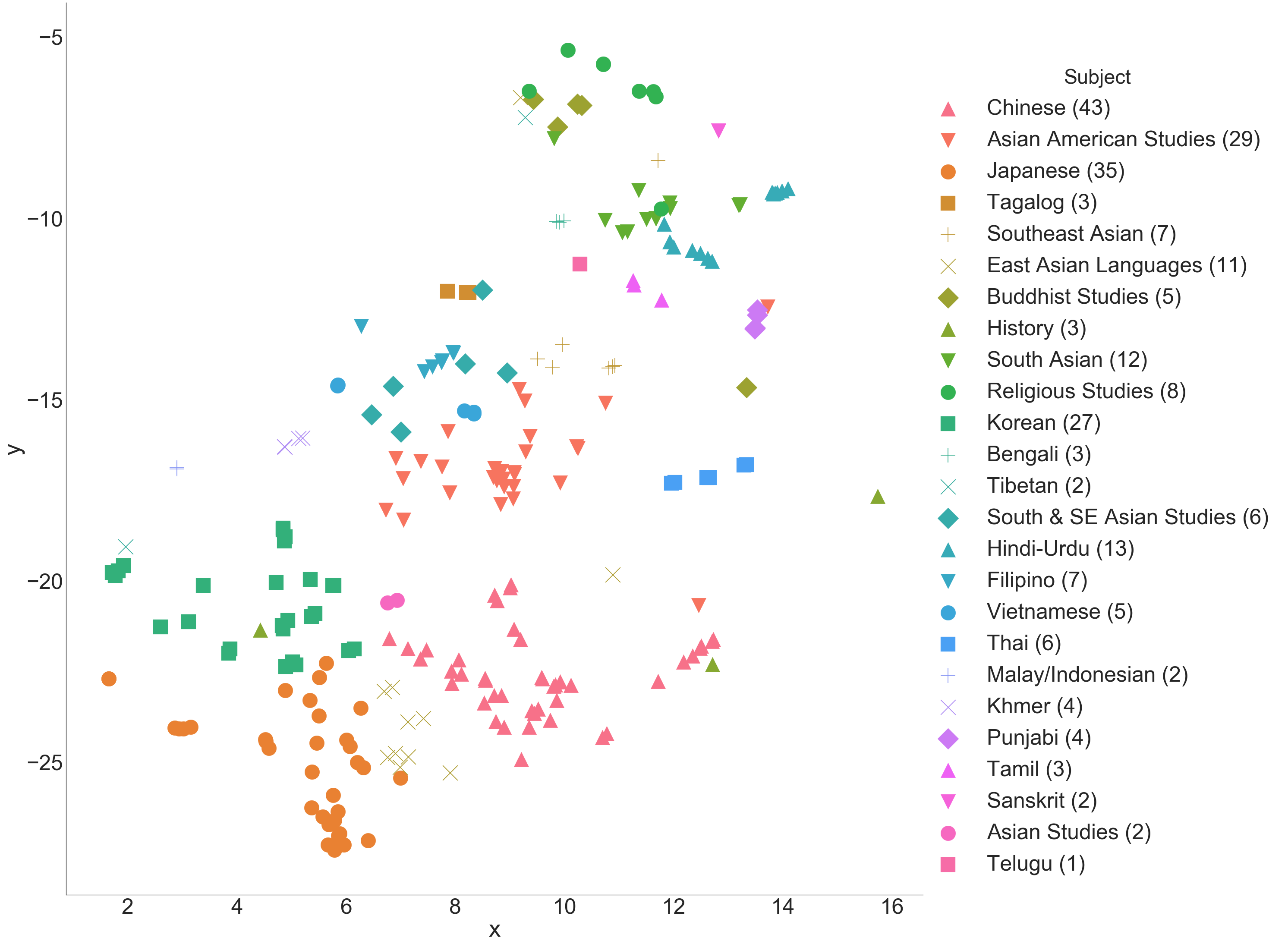}
\caption{Zoom-in of Asian Languages \& Culture cluster}
\label{fig:tsne_meso_alc}
\end{figure*}






\end{document}